\documentclass[letterpaper]{article}

\usepackage{epsfig}

\begin{document}

\title{The bistable brain: a neuronal model \\ with {\it symbiotic} interactions}

\author{Ricardo L\'opez-Ruiz$^1$ and Dani\`ele Fournier-Prunaret$^2$}

\maketitle

\begin{center}
$^1$Department of Computer Science and BIFI, \\ 
Universidad de Zaragoza, 50009 Zaragoza, Spain. \\
{\it rilopez@unizar.es}
\end{center}

\begin{center}
$^2$LAAS-CNRS, 7 Avenue du Colonel Roche, and INSA, \\
Universit\'e de Toulouse, 31077 Toulouse Cedex, France. \\
{\it Daniele.Fournier@insa-toulouse.fr}
\end{center}

\begin{abstract}
 In general, the behavior of large and complex aggregates of elementary 
 components can not be understood nor extrapolated from the properties of 
 a few components. The brain is a good example of this type of networked 
 systems where some patterns of behavior are observed independently of 
 the topology and of the number of coupled units.   
 Following this insight, we have studied the dynamics of different 
 aggregates of logistic maps according to a particular {\it symbiotic} coupling 
 scheme that imitates the neuronal excitation coupling. 
 All these aggregates show some common dynamical properties, 
 concretely a bistable behavior that is reported here with a certain detail.
 Thus, the qualitative relationship with neural systems is suggested through
a naive model of many of such networked logistic maps whose behavior 
mimics the waking-sleeping bistability displayed by brain systems.
Due to its relevance, some regions of multistability are determined and 
sketched for all these logistic models.
\end{abstract}

{\bf Classification:}
{07.05.Mh, 05.45.Ra, 05.45.Xt}


\newpage
\section{INTRODUCTION}

The brain is a natural networked system \cite{cajal,llinas03}. 
The understanding of this complex system
is one of the most fascinating scientific tasks today, 
concretely how this set of millions of neurons can 
{\it symbiotically} interact among them to give 
rise to the collective phenomenon of human thinking \cite{sfn}, or, 
in a simpler and more realistic approach, what neural features can make 
possible, for example, the birdsongs \cite{mindlin}.   
Different aspects of neurocomputation take
contact on this problem: how brain stores information and how brain
processes it to take decisions or to create new information.
Other universal properties of this system are more evident.  One of
them is the existence of a regular daily behavior: the sleep-wake
cycle \cite{winfree, bar-yam}. 

\begin{figure}[h]
 \center\includegraphics[height=.1\textheight]{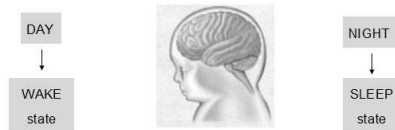}
 \label{2states}
  \caption{Brain bistability provoked by the solar light cycle.}
\end{figure}

The internal circadian rhythm is closely synchronized 
with the cycle of sun light (Fig. \ref{2states}). Roughly speaking and depending on 
the particular species, the brain is awake during the day and it is slept during the
night, or vice versa. All mammals and birds sleep. 
There is not a well established law relating
the size of the animal with the daily time it spends sleeping, but, 
in general, large animals tend to sleep less than small animals (Fig. 2). 
Hence, at first sight, the emergent bistable sleep-wake behavior seems not depend 
on the precise architecture of the brain nor on its size. 
This structural property would mean that, if we represent the brain as 
a complex assembly of units \cite{lopezruiz007,lopezruiz07,lopezruiz09},
the possible bistability, where large groups of neurons 
can show some kind of synchronization, should not depend on the 
topology (structure) nor on the number of nodes (size) of the network (Fig. \ref{waksleep}).

\begin{figure}[h]
 \center\includegraphics[height=.15\textheight]{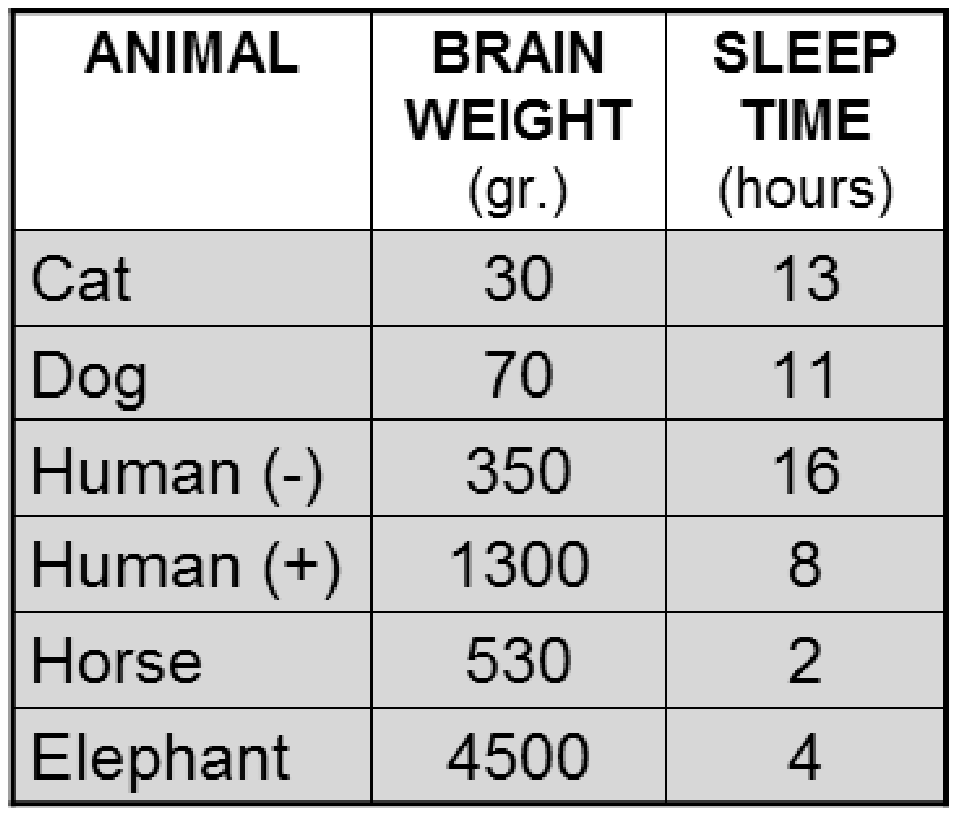}
 \label{table1}
  \caption{Brain size and mean sleeping time for different animals. (Humans(-) represent
  new born humans and Humans(+) represent middle age humans).}
\end{figure}

\begin{figure}[h]
 \center\includegraphics[height=.2\textheight]{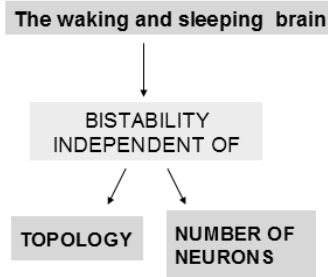}
  \caption{The waking-sleeping brain.}
  \label{waksleep}
\end{figure}

Then, it is essential the type of local dynamics and the excitation/inhibition coupling among 
the nodes that must be implemented in order to get bistability as a possible dynamical state
in a complex network. So, on one side, it has been recently argued \cite{eguiluz} 
that the distribution of functional connections $p(k)$ in the human brain,
where $p(k)$ represents the probability of finding 
an element with $k$ connections to other elements of the network,
follows the same distribution of a scale-free network \cite{barabasi},
that is a power law behavior, $p(k)\sim k^{-\gamma}$, 
with $\gamma$ around $2$. This finding means that there are regions in the
brain that participate in a large number of tasks while most of the regions
are only involved in a tiny fraction of the brain's activities.
On the other side, it has been shown by Kuhn et al. \cite{kuhn} 
the nonlinear processing of synaptic inputs in cortical neurons.
They studied the response of a model neuron with a simultaneous increase of 
excitation and inhibition. They found that the firing rate 
of the model neuron first increases, reaches a maximum, and then decreases at 
higher input rates. Functionally, this means that the firing rate, commonly
assumed to be the carrier of information in the brain, is a non-monotonic
function of balanced input. These findings do not depend on details of the model and, 
hence, they are relevant to cells of other cortical areas as well. 

\begin{figure}[h]
 \centering\includegraphics[height=.1\textheight]{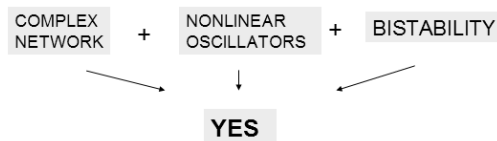}
  \caption{QUESTION: Is it possible to implement some kind of coupling and 
  nonlinear dynamics in each node of a complex network in order to get bistability?. 
  ANSWER: Yes.}
  \label{yes}
\end{figure}

Putting together all these facts, we arrive to the central question that we want
to bring to the reader: Is it possible to reproduce the bistability in a complex
network independently of the topology and of the number of nodes?. The answer
is 'Yes' (Fig. \ref{yes}). What kind of local dynamics and coupling among nodes 
must be implemented in order to get this behavior?. In Section \ref{sec-few}, 
we give different possible strategies for the coupling and the local dynamics 
which should be implemented in a few coupled functional units in order
to get a bistable behavior. Then, in Section \ref{sec-many}, the same model 
is exported on a many units network where the desired bistability is also retained. 
In view of the results, the possibility of constructivism in the world of complex 
systems in general, and in the neural networks in particular, 
is suggested in the Conclusion.

\section{MODELS OF A FEW COUPLED \\ FUNCTIONAL UNITS}
\label{sec-few}

\subsection{General model}

Our approach considers the so called 
{\it functional unit}, i.e. a neuron or group of neurons (voxels), as a discrete 
nonlinear oscillator with two possible states: active (meaning one type of activity) 
or not (meaning other type of activity).
Hence, in this naive vision of the brain as a networked system, 
if $x_n^i$, with $0<x_n^i<1$, represents a measurement of the $ith$ 
functional unit activity at time $n$, 
it can be reasonable to take the most elemental local nonlinearity, 
for instance, a logistic evolution \cite{may}, which presents a quadratic term, 
as a first toy-model for the local neuronal activity:
\begin{equation}
x^i_{n+1} = \bar p_i\;x^i_n(1-x^i_n).
\label{eq0}
\end{equation}

\begin{figure}[h]
 \centering\includegraphics[height=.05\textheight]{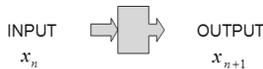}
  \caption{Discrete nonlinear model for the local evolution of a functional unit.}
  \label{1-osc}
\end{figure}

It presents only one stable state for each $\bar p_i$. Then, there is no
bistability in the basic component of our models. For $0<\bar p_i<1$,
the dynamics dissipates to zero, $x_n^i=0$, then it can represent the
functional unit with no activity. For $1<\bar p_i<4$, the dynamics is
non null and it would represent an active functional unit. 

We can suppose that this local parameter $\bar p_i$ is controlled by the signals of 
neighbor units, simulating in some way the effect of the synapses among neurons.
Excitatory and inhibitory synaptic 
couplings have been shown to be determinant on the synchronization of neuronal firing. 
For instance, facilitatory connections are important to explain the neural 
mechanisms that make possible the object representation by synchronization
in the visual cortex \cite{eckhorn}.
While it seems clear that excitatory ({\it symbiotic}) coupling can lead 
to synchronization, frequently inhibition rather than excitation synchronizes 
firing \cite{abbott}. The importance of these two kinds of coupling mechanisms
has also been studied for other types of neurons, v.g., motor neurons \cite{koening}.

If a neuron unit simultaneously processes a plurality of binary input signals,
we can think that this local information processing is reflected by the 
parameter $\bar p_i$. The functional
dependence of this local coupling on the neighbor states is essential
in order to get a good brain-like behavior (i.e., as far as the
bistability of the sleep-wake cycle is concerned) of the network. 
As a first approach, we can take $\bar p_i$ as a linear 
function depending on the actual mean value, $X_n^i$, of the neighboring 
signal activity and expanding the interval $(0,4)$ in the form:
\begin{eqnarray}
\bar p_i & = & p_i\;(3X_n^i+1), \;\;\;\;\;\;\; (excitation \;\; coupling) \label{eq1}\\
& {or} & \nonumber \\
\bar p_i & = & p_i\;(-3X_n^i+4), \;\;\;\; (inhibition \;\; coupling) \label{eq11}
\end{eqnarray}
with 
\begin{equation}
X_n^i={1\over N_i}\sum_{j=1}^{N_i}x_n^j.
\label{eq2}
\end{equation}
$N_i$ is the number of neighbors of the $ith$ functional unit, and
$p_i$, which gives us an idea of the interaction of the functional unit
with its first-neighbor functional units, is the control parameter.
This parameter runs in the range $0<p_i<p_{max}$, where $p_{max}\ge 1$. 
When $p_i=p$ for all $i$, the dynamical behavior of these networks 
with the excitation type coupling \cite{lopezruiz007,lopezruiz07} 
presents an attractive global null configuration 
that has been identified as the {\it turned off}
state of the network.  Also they show a completely synchronized
non-null stable configuration that represents the {\it turned on}
state of the network. Moreover, a robust bistability between these two perfect 
synchronized states is found in that particular model. For different models with 
a few coupled functional units we sketch in the next subsections the regions 
where they present a bistable behavior. The details of the complete 
unfolding \cite{mira} of these dynamical systems can be found in the 
references \cite{lopezruiz04,lopezruiz05,fournier06}.

\subsection{Models of two functional units}

Let us start with the simplest case of two interconnected $(x_n,y_n)$
functional units. Three different combinations of couplings are possible:
$(excitation, excitation)$, $(excitation, inhibition)$ and 
$(inhibition, inhibition)$. 

\begin{figure}[h]
  \centering\includegraphics[height=.05\textheight]{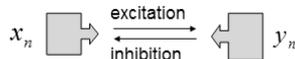}
  \caption{Two functional coupled units.}
\end{figure}

As we are concerned with interactions with a certain degree of symbiosis between 
the coupled units, the regions of parameter space where bistability is found for 
the first two cases are presented here. Experimental systems where similar couplings
have been found or implemented for two cells systems have been reported in the 
literature \cite{graham2007,mihailovic2012}.

\subsubsection{Model with mutual excitation}

\begin{figure}[h]
  \centering\includegraphics[height=.18\textheight]{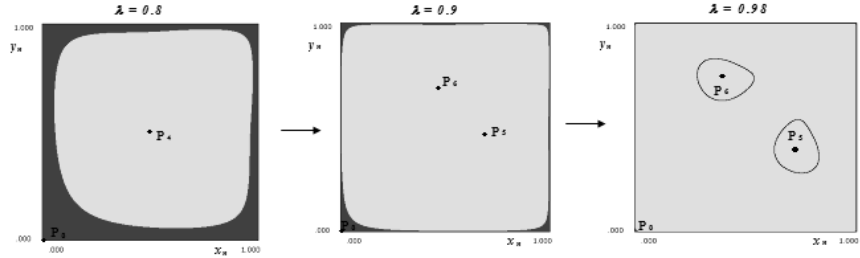}
  \caption{Bistability in $2$ functional units with excitation type coupling ($p=\lambda$).}
  \label{2func+}
\end{figure}

The dynamics of the $(excitation,excitation)$ case \cite{lopezruiz04}
is given by the coupled equations:
\begin{eqnarray}
x_{n+1} & = & p \;(3y_n+1)x_n(1-x_n),\\
y_{n+1} & = & p \;(3x_n+1)y_n(1-y_n).
\label{2-osc}
\end{eqnarray}
The regions of the parameter space (Fig. \ref{2func+})
where we can find bistability are:

\begin{itemize}
\item For $0.75<p<0.86$, the synchronized state, $x_+=(\bar x, \bar x)=P_4$,
with $\bar x={1\over 3}\{1+(4-{3\over p})^{1\over 2}\}$, which arises
from a saddle-node bifurcation for the critical value $p=0.75$, is a
stable {\it turned on} state.  This state coexists with the 
{\it turned off} state $x_\theta=0$.
The system presents now bistability and depending on the initial
conditions, the final state can be $x_\theta$ or $x_+$. Switching on
the system from $x_\theta$ requires a level of noise in both functional units
sufficient to render the activity on the basin of attraction of $x_+$.
On the contrary, switching off the two functional units network can be done, for
instance, by making zero the activity of one functional unit, or by doing the
coupling $p$ lower than $0.75$.
\item For $0.86<p<0.95$, the active state of the network is now a
period-$2$ oscillation, namely the period-$2$ cycle $(P_5,P_6)$ in Fig. \ref{2func+}. 
This new dynamical state bifurcates from $x_+$ for $p=0.86$.  
A smaller noise is necessary to activate the system
from $x_\theta$.  Making zero the activity of one functional unit continues to
be a good strategy to turn off the network.
\item For $0.95<p<1$, the active state acquires a new frequency and
presents quasiperiodicity (the invariant closed curves of Fig. \ref{2func+}).  
It is still possible to switch off the
network by putting to zero one of the functional units.
\end{itemize}

\subsubsection{Model with excitation + inhibition}

\begin{figure}[h]
  \centering\includegraphics[height=.18\textheight]{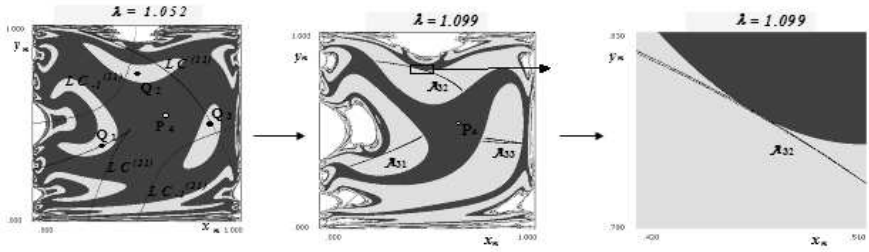}
  \caption{Bistability in $2$ functional units with excitation+inhibition 
            type coupling ($p=\lambda$).}
  \label{2func-}
\end{figure}

The dynamics of the $(excitation,inhibition)$ case \cite{lopezruiz05}
is given by the coupled equations:
\begin{eqnarray}
x_{n+1} & = & p\;(3y_n+1)x_n(1-x_n),\\
y_{n+1} & = & p\;(-3x_n+4)y_n(1-y_n).
\label{2-oscc}
\end{eqnarray}
The regions of the parameter space (Fig. \ref{2func-})
where we can find bistability are:

\begin{itemize}
\item For $1.051<p<1.0851$, a stable period three cycle $(Q_1,Q_2,Q_3)$ appears 
in the system. It  coexists with the fixed point $P_4$. When $p$ is increased, 
a period-doubling cascade takes place and generates successive cycles of higher 
periods $3\cdot 2^n$. The system presents bistability. Depending on the initial conditions, 
both populations $(x_n,y_n)$ 
oscillate in a periodic orbit or, alternatively, settle down in the fixed point.
The borders between the two basins are complex. 
\item For $1.0851<p<1.0997$, an aperiodic dynamics is possible. 
The period-doubling cascade has finally given birth to an order three cyclic chaotic 
band(s) $(A_{31},A_{32},A_{33})$. The system can now present an irregular oscillation 
besides the stable equilibrium with final fixed populations. The two basins are now fractal.
\end{itemize}

\subsection{Models of three functional units}

Following the strategy given by relation (\ref{eq1}-\ref{eq11}) several models with 
three functional units can be established. We have studied in some detail three of 
them \cite{lopezruiz09,fournier06} and their bistable behavior is reported here.

\subsubsection{Model with local mutual excitation}

Let us start with the case of three alternatively interconnected $(x_n,y_n,z_n)$
functional units under a mutual excitation scheme. 

\begin{figure}[h]
  \centering\includegraphics[height=.1\textheight]{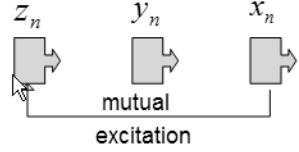}
  \caption{Three alternatively coupled functional units under the excitation scheme.}
\end{figure}

Then the dynamics of the system is given by the coupled equations:
\begin{eqnarray}
x_{n+1} & = & p \;(3y_n+1)x_n(1-x_n), \\
y_{n+1} & = & p \;(3z_n+1)y_n(1-y_n), \\
z_{n+1} & = & p \;(3x_n+1)z_n(1-z_n).
\label{3-osc+}
\end{eqnarray}

\begin{figure}[h]
  \centering\includegraphics[height=.25\textheight]{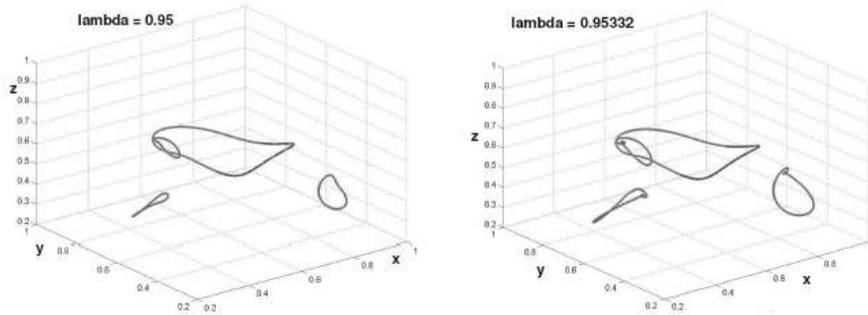}
  \caption{Bistability in $3$ functional units with local excitation type coupling ($p=\lambda$).}
  \label{3func+}
\end{figure}

The regions of the parameter space where we have found bistability are:

\begin{itemize}
\item For $0.93310<p<0.95334$, a big invariant closed curve (ICC) $C1$ coexists 
with a period-$3$ orbit that bifurcates, first to an order 
$3$-cyclic ICC (Fig. \ref{3func+}), and finally to an order-$3$ 
weakly chaotic ring (WCR) before disappearing.
\item For $0.98418<p<0.98763$, the ICC $C1$ coexists with another ICC $C2$ 
(see Ref. \cite{fournier06}) that becomes chaotic, by following a period doubling
cascade of tori, before disappearing. 
\item For $1.00360<p<1.00402$, the ICC $C1$ coexists with a high period orbit
that gives rise to an ICC $C3$. This ICC also becomes a chaotic band
(see Ref. \cite{fournier06}) by following a period doubling
cascade of tori before disappearing. 
\end{itemize}

\subsubsection{Model with global mutual excitation}

We expose now the case of three globally interconnected $(x_n,y_n,z_n)$
functional units under a mutual excitation scheme. 

Then the dynamics of the system is given by the coupled equations:
\begin{eqnarray}
x_{n+1} & = & p\; (x_n+y_n+z_n+1)x_n(1-x_n), \\
y_{n+1} & = & p\; (x_n+y_n+z_n+1)y_n(1-y_n), \\
z_{n+1} & = & p\; (x_n+y_n+z_n+1)z_n(1-z_n).
\label{3-oscil}
\end{eqnarray}

\begin{figure}[h]
  \centering\includegraphics[height=.13\textheight]{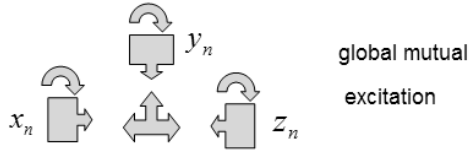}
  \caption{Three globally coupled functional units under the excitation scheme.}
\end{figure}

\begin{figure}[h]
  \centering\includegraphics[height=.25\textheight]{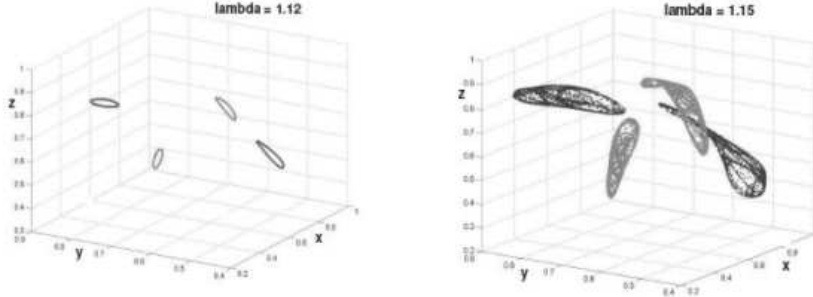}
  \caption{Bistability in $3$ functional units with global excitation type coupling ($p=\lambda$).}
  \label{3-oscc}
\end{figure}

For the whole range of the parameter, $0<p<1.17$, bistability is present in this system:

\begin{itemize}
\item   Firstly, two order-$2$ cyclic ICC coexist before becoming two order-$2$ 
cyclic chaotic attractors (Fig. \ref{3-oscc}) by contact bifurcations of heteroclinic type.
Finally the two chaotic attractors become a single one before disappearing.
\end{itemize}

\subsubsection{Model with partial mutual excitation}

\begin{figure}[h]
 \center\includegraphics[height=.1\textheight]{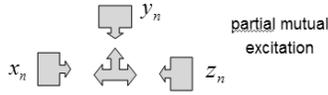}
  \caption{Three partially coupled functional units under the excitation scheme.}
  \label{fig-3}
\end{figure}

The new case \cite{lopezruiz09} of three partially interconnected $(x_n,y_n,z_n)$
functional units under a mutual excitation scheme is represented in Fig. \ref{fig-3}. 

The dynamics of the system is given by the coupled equations:
\begin{eqnarray}
x_{n+1} & = & p\; (3(y_n+z_n)/2+1)x_n(1-x_n), \\
y_{n+1} & = & p\; (3(x_n+z_n)/2+1)y_n(1-y_n), \\
z_{n+1} & = & p\; (3(x_n+y_n)/2+1)z_n(1-z_n).
\label{3-osc-1}
\end{eqnarray}

\begin{figure}[] 
 \center\includegraphics[height=.28\textheight]{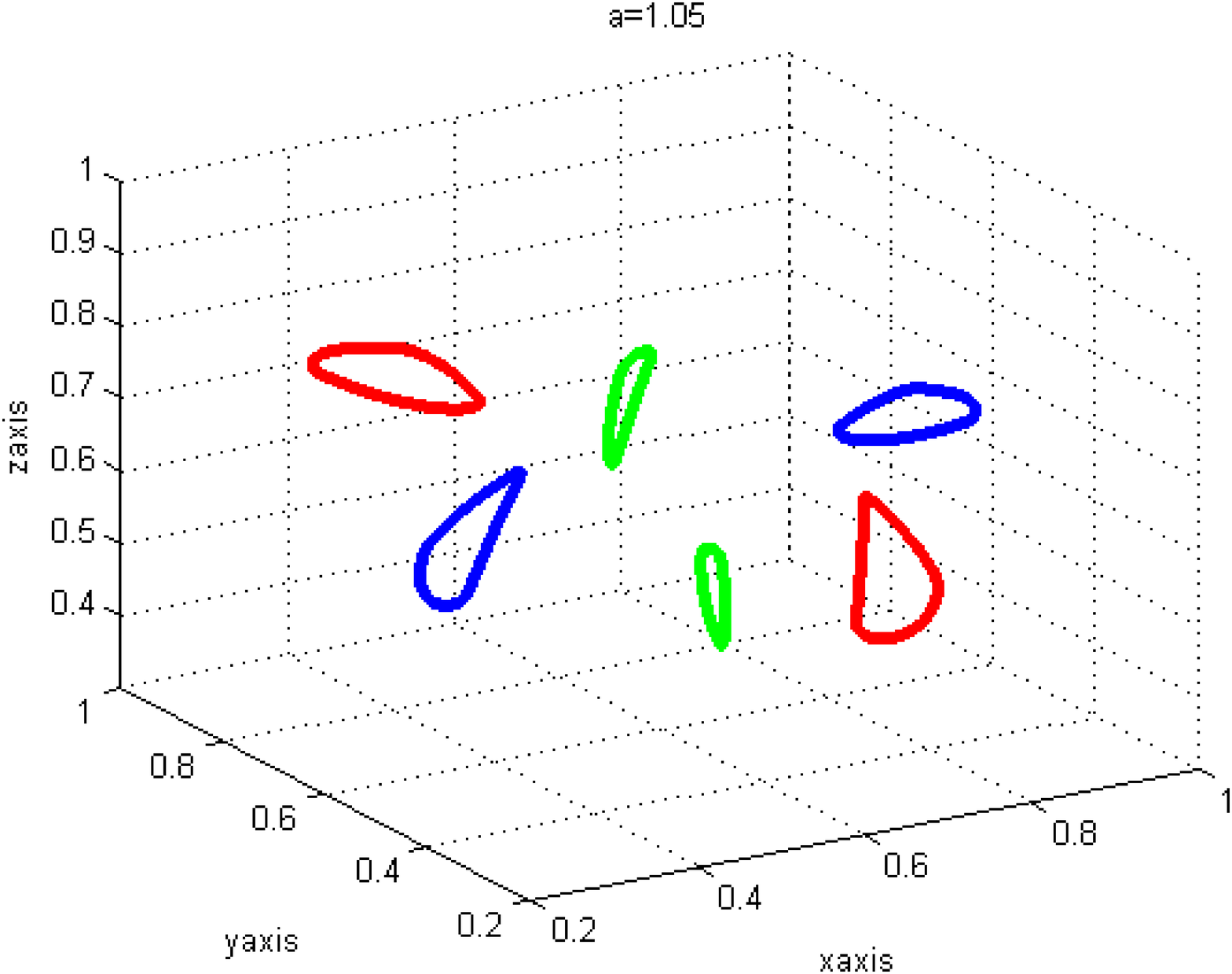}
  \caption{Multistability in $3$ functional units with partial excitation type coupling.
  The system presents three order-$2$ ICC for $p=a=1.05$.}
  \label{fig6}
\end{figure}

\begin{figure}[h]
 \center\includegraphics[height=.28\textheight]{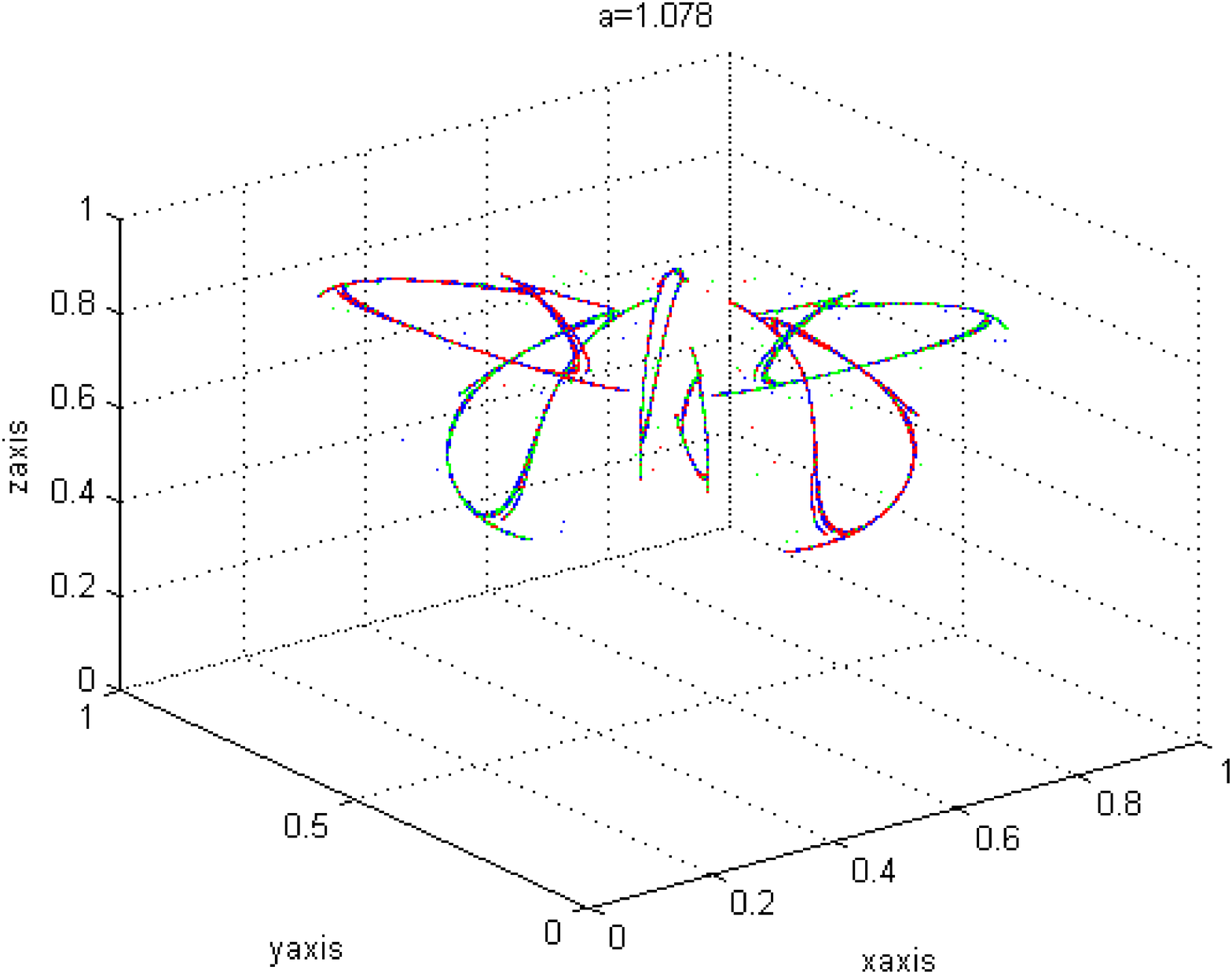}
  \caption{Multistability of three chaotic cyclic attractors of order $2$ for $p=a=1.078$.}
  \label{fig7}
\end{figure}

\begin{figure}[]
 \center\includegraphics[height=.28\textheight]{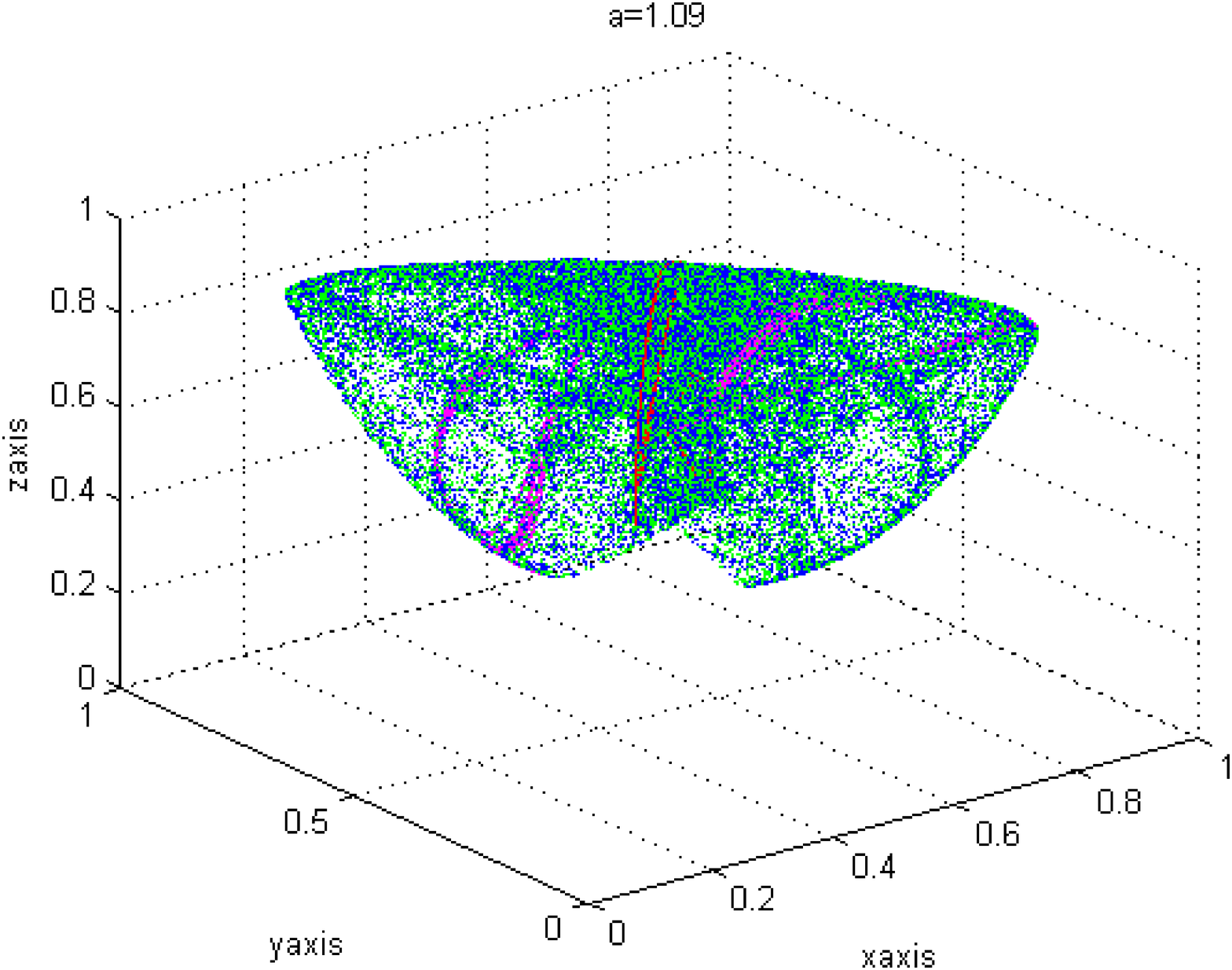}
  \caption{An unique chaotic attractor for $p=a=1.09$.}
  \label{fig8}
\end{figure}

\begin{figure}[h]
 \center\includegraphics[height=.28\textheight]{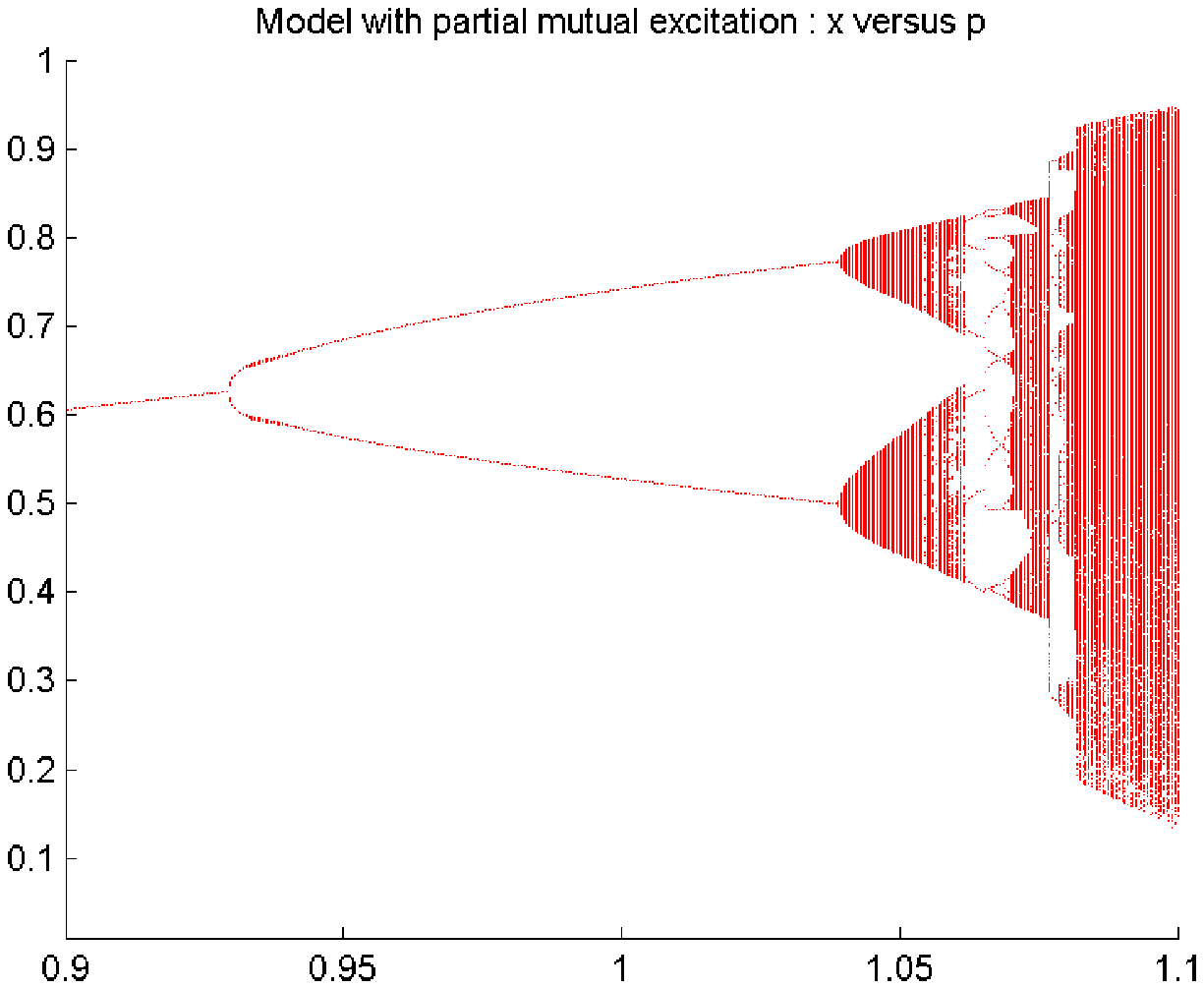}
  \caption{Bifurcation diagram projected in the $x$ coordinate for the initial conditions 
  $(x_0=0.37,y_0=0.36,z_0=0.33)$ as a function of the parameter $p$.}
  \label{fig9}
\end{figure}

The rough inspection of this system puts in evidence 
the existence of different regions of multistability 
in the parameter space. These are:

\begin{itemize}
\item For $0.93<p<1.04$, there is coexistence among
three cycles of period-$2$.
\item For $1.04<p<1.06$, the three cycles bifurcate giving rise
to three order-$2$ ICC (Fig. \ref{fig6}).
\item For $1.06<p<1.08$, the system can present three mode-locked periodic orbits, 
each one with period multiple of $6$, displaying period doubling cascades giving rise to
three chaotic cyclic attractors. Three chaotic cyclic attractors 
of order $2$ are also possible in this region (Fig. \ref{fig7}). 
\item For $p>1.08$, the chaotic cyclic attractors collapse in an unique chaotic 
attractor (Fig. \ref{fig8}).
\end{itemize}
Also, other multistable situations can be found for some particular values of the parameter $p$
in the former intervals, such as it can be seen in Fig. \ref{fig9}, where the $x$-projection of 
a generic bifurcation diagram is plotted.
Similar diagrams are found for the $y$- and $z$-projections. 

\section{MODEL OF MANY COUPLED \\ FUNCTIONAL UNITS}
\label{sec-many}

\subsection{The model}

The complete synchronization \cite{lopezruiz91,boccaletti02} of the 
network means that $x_n^i=x_n$ for all $i$ , with $i=1,2,\ldots,N$ 
and $N\gg 1$ \cite{jalan03,oprisan09}.  
In this regime, we also have $X_n^i=x_n$. The time
evolution of the network \cite {lopezruiz07} on the synchronization manifold 
is then given by the cubic mapping:
\begin{equation}
x_{n+1} = p\;(3x_n+1)\;x_n(1-x_n).
\label{eq4}
\end{equation}  
The fixed points of this system are found by solving $x_{n+1}=x_n$.
The solutions are $x_\theta=0$ and $x_\pm={1\over 3}\{1\pm(4-{3\over
p})^{1\over 2}\}$.  The first state $x_\theta$ is stable for $0<p<1$
and $x_\pm$ take birth after a saddle-node bifurcation for
$p=p_0=0.75$. The node $x_+$ is stable for $0.75<p<1.157$ and the
saddle $x_-$ is unstable.  Therefore bistability between the states
\begin{eqnarray}
x_n^i = x_\theta, & \forall i \longrightarrow & TURNED\;\;\; OFF\;\;\; STATE, \\
x_n^i = x_+, & \forall i \longrightarrow& TURNED\;\;\; ON\;\;\; STATE, 
\label{eq5}
\end{eqnarray}
seems to be also possible for $p>p_0=0.75$ in the case of many interacting units.  
But stability on the synchronization manifold does not imply
the global stability of it.  Small transverse perturbations to this manifold
can make unstable the synchronized states.  Let us suppose then a
general local perturbation $\delta x_n^i$ of the element activity,
\begin{equation} 
x_n^i=x_*+\delta x_n^i,
\label{eq6}
\end{equation}
with $x_*$ representing a synchronized state, $x_{\theta}$ or $x_+$.  
We define the perturbation of the local mean-field as
\begin{equation} 
\delta X_n^i={3\over N_i}\sum_{j=1}^{N_i}\delta x_n^j.
\label{eq7} 
\end{equation}
If these expressions are introduced into equation (\ref{eq0}),
the time evolution of the local perturbations are found:
\begin{equation}
\delta x_{n+1}^i=p\,(3x_*+1)(1-2x_*)\delta x_n^i+p\,x_*(1-x_*)\delta X_n^i.
\label{eq8}
\end{equation} 
The dynamics for the local mean-field perturbation is derived by
substituting this last expression in relation (\ref{eq7}). We obtain:
\begin{equation}
\delta X_{n+1}^i=p\,(3x_*+1)(1-2x_*)\delta X_n^i+
3p\,x_*(1-x_*){1\over N_i}\sum_{j=1}^{N_i}\delta X_n^j.
\label{eq9}
\end{equation}
We express now the local mean-field perturbations of the first-neighbors 
as function of the local mean-field perturbation $\delta X_n^i$ 
by defining the local operational quantity $\sigma_i^n$,
\begin{equation}
{1\over N_i}\sum_{j=1}^{N_i}\delta X_n^j=\sigma_n^i\;\delta X_n^i,
\label{eq10}
\end{equation}
which is determined by the dynamics itself.  If we
put together the equations (\ref{eq8}-\ref{eq9}), the linear stability
of the synchronized states holds as follows:
\begin{equation}
\left(\begin{array}{c} \delta x_{n+1}^i \\ \delta X_{n+1}^i \end{array} \right) =
\left(\begin{array}{cc} 
p\,(3x_*+1)(1-2x_*) & p\,x_*(1-x_*) \\
0 & p\,(3x_*+1)(1-2x_*)+ 3p\,\sigma_n^i\,x_*(1-x_*)
\end{array} \right)
\left(\begin{array}{c} \delta x_{n}^i \\ \delta X_{n}^i \end{array} \right).
\label{eq111}
\end{equation}
Let us observe that the only dependency on the network topology is
included in the quantity $\sigma_n^i$. The rest of the stability
matrix is the same for all the nodes and therefore it is independent
of the local and global network organization.

The turned off state is $x_*=x_\theta=0$. The eigenvalues of the
stability matrix are in this case $\lambda_1=\lambda_2=p$. Then, this
state is an attractive state in the interval $0<p<1$.  It loses
stability for $p=1$, then the highest value $p_f$ of the parameter $p$ 
where bistability is still possible satisfies $p_f\leq 1$.

\begin{figure}[]
\center\includegraphics[height=.40\textheight,angle=-90]{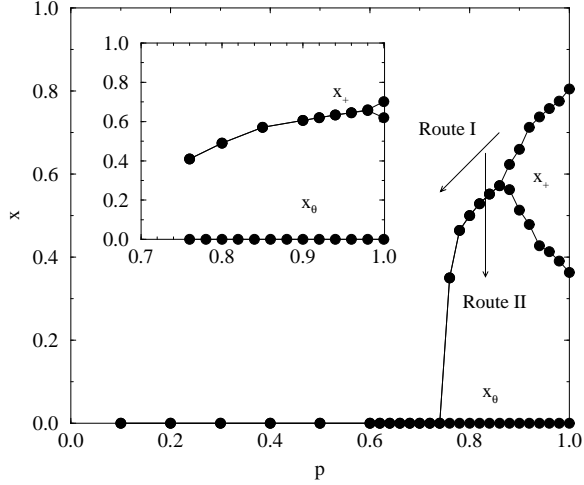}
\caption{Stable states $(x_{\theta},x_+)$ of the network for
$0<p<1$.  Let us observe the two zones of bistability: $p_0<p<p_c$ and
$p_c<p<p_f$.  The main figure corresponds to a scale free network made
up of $N=10^4$ elements: $p_0=0.75$, $p_c=0.87\pm 0.01$ and $p_f=1$. The
inset shows the same graph but in an all-to-all network of the same
size: $p_0=0.75$, $p_c=p_f=1$. Initial conditions for the $x_i$'s were
drawn from a uniform probability distribution in the interval $(0,1)$.}
\label{fig1}
\end{figure}

The turned on state $x_+$ verifies $x_*=x_+={1\over 3}\{1+(4-{3\over
p})^{1\over 2}\}$.  If we suppose $\sigma_n^i=\sigma$, the eigenvalues
of the stability matrix are $\lambda_1=2-2p-p(4-{3\over p})^{1\over
2}$ and $\lambda_2=\lambda_1+{\sigma\over 3}(3-2p+p(4-{3\over
p})^{1\over 2})$.  Let us observe that $\lambda_1=-1$ for $p=1$. This
implies that the parameter $p_c$ for which the synchronized state
$x_+$ looses stability verifies $p_c\leq 1$.  Depending on the sign of
$\sigma$, we can distinguish two cases in the behavior of $p_c$:
\begin{itemize}
\item If $0<\sigma<1$, we find that $\mid\lambda_2\mid<1$. Then $x_+$
bifurcates through a global flip bifurcation for $p=p_c=1$. In
this case, the bifurcation of the synchronized state $x_+$ for $p_c=1$
coincides with the loss of the network bistability for $p_f=1$. Hence
$p_c=p_f=1$ for this kind of networks, and the bistability holds
between $x_\theta$ and $x_+$ in the parameter interval
$p_0=0.75<p<p_c=p_f=1$.  As an example, an all-to-all network shows
this behavior because $\sigma=1$. This is represented in the inset 
of Fig. \ref{fig1}.
\item If $-1<\sigma<0$, then $\lambda_2=-1$ is obtained for a $p=p_c$
smaller than $1$. Therefore it is now possible to obtain an active
state different from $x_+$ in the interval $p_c<p<p_f$.  For instance,
simulations show that the global flip bifurcation of the synchronized
state for a scale free network occurs for $p_c=0.87\pm 0.01$.  A value 
of $p=0.866$ is
obtained from the stability matrix by taking $\sigma=-1$.  For this
particular network it is also found that $p_f=1$. Then, bistability is
possible in the range $p_0=0.75<p<p_f=1$ for this kind of
configuration.  But now an active state with different dynamical
regimes is observed in the interval $p_c=0.87<p<p_f=1$. If we identify
the capacity of information storing with the possibility of the system
to access to complex dynamical states, then, we could assert, in this
sense, that a scale free network has the possibility of storing more
elaborated information in the bistable region that an all-to-all
network.
\end{itemize}

Let us note that $\sigma$ also indicates a different behavior of local
dissipation, as expression (\ref{eq10}) suggests.  A positive $\sigma$
means a local in-phase oscillation of the node signal and mean-field
perturbations. A negative $\sigma$ is meaning a local out of phase
oscillation between those signal perturbations. Hence, $\sigma$ also brings
some kind of structural network information.  In all the cases the
stability loss of the completely synchronized state is mediated by a
global flip bifurcation. The new dynamical state arising from that
active state for $p=p_c$ is a periodic pattern with a local period-$2$
oscillation. The increasing of the coupling parameter monitors other
global bifurcations that can lead the system towards a pattern of local
chaotic oscillations.

\subsection{Transition between On-Off States}

We proceed now to show the different strategies for switching on and
off a random scale free network.  The choice of this network is
suggested by the recent work \cite{eguiluz} 
on the connections distribution among
functional units in the brain.  They find it to be a power-law
distribution.  Following this insight \cite{lopezruiz07}, 
a scale-free network following 
the Barab\'asi-Albert (BA) recipe \cite{barabasi} is generated.  
In this model, starting from a set of $m_0$ nodes, one
preferentially attaches each time step a newly introduced node to $m$
older nodes. The procedure is repeated $N-m_0$ times and a network of
size $N$ with a power law degree distribution $P(k)\sim k^{-\gamma}$
with $\gamma=3$ and average connectivity $\langle k \rangle=2m$ builds
up. This network is a clear example of a highly heterogenous network,
in that the degree distribution has unbounded fluctuations when
$N\rightarrow\infty$. The exponent reported for the brain functional
network has $\gamma < 3$. However, studies of percolation and epidemic
spreading \cite{callaway,dorogovtsev03} on top of scale-free networks have
shown that the results obtained for $\gamma=3$ are consistent with
those corresponding to lower values of $\gamma$, with $\gamma > 2$. 
As explained before, network bistability between
the active and non active states is here possible in the interval
$p_0=0.75<p<p_f=1$ (Fig. \ref{fig1}).

\subsubsection*{Switching off the network}

\begin{figure}[]
\center\includegraphics[height=.40\textheight,angle=-90]{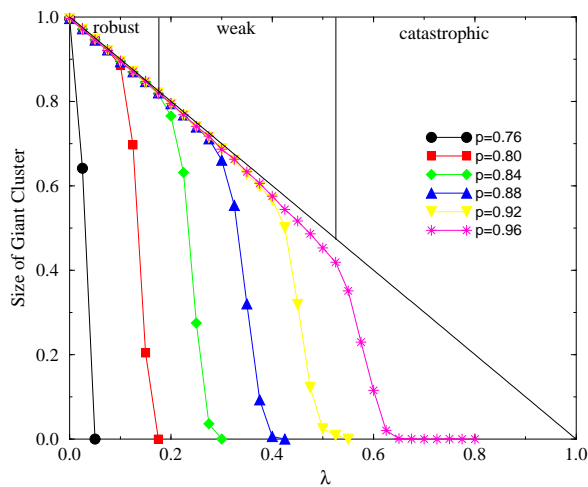}
\caption{Turning off a scale free network. Three different phases
in the behavior of the giant cluster size versus $\lambda$ (fraction
of switched off nodes) are observed. These three phases are illustrated 
for $p=0.96$: the robust phase, the weak phase and the catastrophic 
phase (see the text). Other network parameters are as those of Fig. \ref{fig1}.}
\label{fig2}
\end{figure}

Two different strategies \cite{lopezruiz07} can be followed to carry 
the network from the active state to that with no activity (Fig. \ref{fig1}).
\begin{itemize}
\item {\it Route I}: By doing the coupling $p$ lower than
$p_0$.  This is the easiest and more natural way of performing such an
operation.  In our naive picture of a brain-like system, 
it could represent the decrease 
(or increase, it depends on the specific function)
of the synaptic substances that provokes the transition
from the awake to the sleep state. The flux of these chemical
activators is controlled by the internal circadian clock, which is
present in all animals, and which seems to be the result of living
during millions of years under the day/night cycle.
\item {\it Route II}: By switching off a critical fraction of functional units
for a fixed $p$. Evidently, at first sight, this strategy has no relation with 
the behavior of a real brain-like system. 
Thus, this is done by looking over all the elements of the
network, and considering that the element
activity is set to zero with probability $\lambda$ 
(which implies that on average $\lambda N$
elements are reset to zero). The result of this operation is shown in
Fig. \ref{fig2}. Here, it is plotted for different $p$'s the relative size of
the biggest (giant) cluster of connected active nodes in the network
versus $\lambda$. Note that this procedure does not take into account
the existence of connectivity classes, but all nodes are equally
treated. The procedure is thus equivalent to simulations of random
failure in percolation studies \cite{callaway}. The strategy in which 
highly connected functional units are first put to zero is more aggressive 
and leads to quite different results. 
\end{itemize}

Each curve presents three different zones depending on $\lambda$:
\begin{quote}
\item - the {\it robust phase}: For small $\lambda$, the network is
stable and only those states put to zero have no activity.  There is a
linear dependence on the giant cluster size with $\lambda$.  In this
stage, the switched off nodes do not have the capacity to transmit its
actual state to its active neighbors.
\item - the {\it weak phase}: For an intermediate $\lambda$, the nodes
with null activity can influence its neighborhood and switch off some
of them. The linearity between the size of the giant cluster and
$\lambda$ shows a higher absolute value of the slope than in the
robust zone.
\item - the {\it catastrophic phase}: When a critical $\lambda_c$ is
reached, the system undergoes a crisis. The sudden drop in this zone
means that a small increase of the non active nodes leads the system to
a catastrophe; that is, the null activity is propagated through all
the network and it becomes completely down.
\end{quote}

It is worth noticing that when the system is outside the bistability
region for $p>1$, the catastrophic phase does not take place. Instead,
the turned off nodes do not spread its dynamical state and the
neighboring nodes do not die out. This is because the dynamics of
an isolated node is self-sustained when $p>1$. Consequently, it is observed
that the network breaks down in many small clusters and the transition
resembles that of percolation in scale free nets \cite{callaway,vazquez03}.

\subsubsection*{Switching on the network}

\begin{figure}[]
\center\includegraphics[height=.40\textheight,angle=-90]{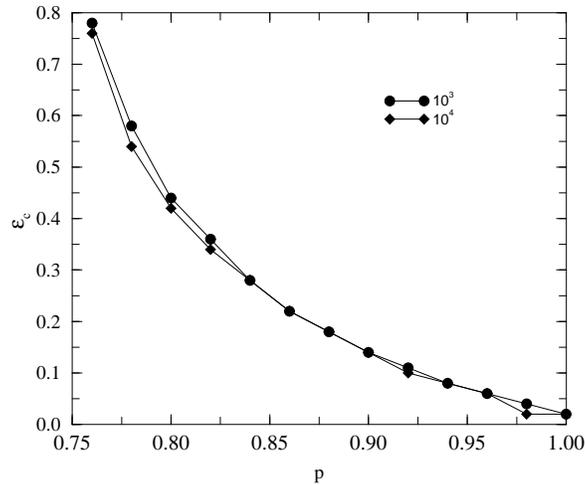}
\caption{Turning on a scale free network. For a fixed $p$, a noisy
signal randomly distributed in the interval $(0,\epsilon)$ is
assigned to every node. When $\epsilon$ reaches the
critical level $\epsilon_c$ the network becomes switched
on. Other network parameters are as those of Fig. \ref{fig1}.}
\label{fig3}
\end{figure}

Two equivalent strategies \cite{lopezruiz07} can be followed for the case
of turning on the network (Fig. \ref{fig3}): 
\begin{itemize}
\item (I) For a fixed $p$, we can increase
the maximum value $\epsilon$ of the noisy signal, which is randomly
distributed in the interval $(0,\epsilon)$ over the whole system.
When $\epsilon$ attains a critical value $\epsilon_c$, the noisy
configuration can leave the basin of attraction of $x_{\theta}$,
whose boundary seems to have the form in phase space 
of a ``hollow cane'' around it, and then 
the network rapidly evolves toward the turned on state; 
\item (II) If this operation is executed by letting $\epsilon$ to be 
fixed and by increasing the coupling parameter $p$, the final result 
of switching on the network is reached when $p$ takes the value for which
$\epsilon=\epsilon_c$. The final result is identical in both cases.
\end{itemize}

Let us remark that the strategy equivalent to the former Route II,
that is, the switching on of a critical fraction of functional units,
is not possible in this case. It is a consequence of the fact that a
switched off functional unit can not be excited by its neighbors and it will
maintain indefinitely the same dynamical state ($x_i=0$).

Finally, let us note that, from Fig. \ref{fig3}, a bigger $p$ requires
a smaller $\epsilon_c$ to switch on the net. Observe that this behavior 
could be interpreted in our approach as the smaller level of noise 
that is needed for awaking a brain-like system that it is departing from 
the sleeping state.

\section{CONCLUSION}

One of the more challenging problems in nonlinear science is the goal 
of understanding the properties of neuronal circuits \cite{varona}. 
Synchrony and multistability are two important dynamical behaviors found
in those circuits \cite{borgers03,hansel03,buzsaki04}.  
In this work, different coupling schemes for networks with local logistic
dynamics are proposed \cite{lopezruiz007}. It is observed that these types 
of couplings generate a global bistability between two different 
dynamical states \cite{lopezruiz07,lopezruiz09}.
This property seems to be topology and size independent. This is a direct
consequence of the local mean-field multiplicative coupling among the
first-neighbors. If a formal and naive relationship is established between 
these two states and the sleep-wake states of a brain, respectively, 
one would be tempted to assert that these types of couplings in a network,
regardless of its simplicity, give us a good qualitative model for
explaining that specific bistability. 
Following this insight, several low-dimensional systems with logistic components  
coupled under these schemes have been presented and the regions where the dynamics
shows bistability have been identified. The extension of this type of coupling 
to a general network with local logistic dynamics has also been achieved.
This system presents global bistability between an active
synchronized state and another synchronized state with no activity.
This property is topology and size independent. This is a direct
consequence of the local mean-field multiplicative coupling among the
first-neighbors. If a formal and naive relationship is established between the
switched off and switched on states of that network, and the
sleep-wake states of a brain, respectively, one would be tempted to
assert that this model, regardless of its simplicity, 
is a good qualitative representation for explaining that specific bistability. 
Furthermore, on more theoretical grounds, the results obtained here point out 
the entangled interplay between topology and function in networked 
systems \cite{strogatz01} where complex structures coexist with nonlinear dynamics.


\end{document}